# Inter-Core Crosstalk in Multicore Fibers: Impact on 56-Gbaud/λ/Core PAM-4 Transmission

Aleksejs Udalcovs[1], Rui Lin[2,3], Oskars Ozolins[1], Lin Gan[3], Lu Zhang[2], Xiaodan Pang[4], Richard Schatz[2], Anders Djupsjöbacka[1], Ming Tang[3], Songnian Fu[3], Deming Liu[3], Weijun Tong[5], Sergei Popov[2], Gunnar Jacobsen[1], and Jiajia Chen[2]

[1] Networking and Transmission Laboratory, RISE Acreo AB, Kista, Sweden, aleksejs.udalcovs@ri.se
[2] Schools of EECS and SCI, KTH Royal Institute of Technology, Kista, Sweden, jiajiac@kth.se
[3] Huazhong University of Science and Technology, Wuhan, China, tangming@hust.edu.cn
[4] Infinera, Stockholm, Sweden, xpang@infinera.com
[5] Yangtze Optical Fiber and Cable Joint Stock Limited Company, China, tongweijun@yofc.com

**Abstract** *We experimentally demonstrate the impact of inter-core crosstalk in multicore fibers on 56-Gbaud PAM-4 signal quality after 2.5-km transmission over a weakly-coupled and uncoupled seven-core fibers, revealing the crosstalk dependence on carrier central wavelength in range of 1540-1560 nm.*

**Introduction**

The towering increase of traffic demand in datacenters imposes the need for solutions that, *on one hand*, are capable of scoping with the constantly growing bandwidth requirements[1] and, *on other hand*, are dealing the bandwidth-density issues[2]. In this regard, a spatial division multiplexing (SDM) is proposed on top of high-speed and low-cost solutions of short-reach interfaces for addressing the bandwidth-density problem and for scalability improvements[2,3]. A SDM can be realized in a single fiber through a separation of either signal's modes, cores or combination of two. High spatial density leads to a potentially high number of parallel transmission lanes, but at the same time induces a strong coupling effect between the spatially-separated channels and, therefore, giving a penalty in terms of transmission performance. Weakly-coupled multicore fibers (MCF) is an option, in which a residual coupling effect is present, but it does not impose additional requirements on signal processing at the receiver[4]. Some of the previous works have been carried out modelling the time[5] and frequency[6] dependencies of the inter-core crosstalk in multicore fibers of different type and structural design. However, the experimental validation for high-speed data transmission (100-Gbps and beyond) has not been done yet. Such analyses would provide essential data for designing a MCF structure with pre-defined characteristics for data transmission.

In this paper, for the first time to the best of our knowledge, we experimentally determine the impact that the time- and frequency dependent inter-core crosstalk in multicore fibers has on 56-Gbaud PAM-4 signals transmitted over 2.5-km fiber links for intra-datacenter networks, which use MCFs with similar geometrical design and fiber parameters (such as attenuation and dispersion) but different inter-core crosstalk. The statistical analysis is employed on more than 1 000 BER measurements to reveal the time- and frequency dependence, which may result in transmission penalty for a deployed link.

**Experimental setup and fibers**

The setup used for this purpose is illustrated in Fig. 1. To generate a PAM-4 signal, we used an approach that could be described as *"attenuate-delay-and-combine"*. At the transmitter, two electrical signals from a pulse pattern generator (PPG), representing a $2^{15}-1$ long pseudorandom binary sequence (PRBS15) at 56-Gbaud, are decorrelated and then passively combined into a single 56-Gbaud PAM-4 signal. This signal is amplified to drive a 40-GHz external Mach-Zehnder modulator (MZM) that is used to modulate optical carrier from an external cavity laser (ECL). Optical carrier's central wavelength is set to 1540 nm, 1552 nm and 1560 nm to cover a major part of the C-band. The modulated optical signal is then amplified by an Erbium doped fiber amplifier (EDFA) to compensate the losses that appear due to 1:8 signal splitting and decorrelation before the spatially distributed channels are coupled into the 2.5-km seven-core fiber via a fan-in device (FI). In this experiment, we used two different types of hexagonal multicore fibers: a weakly-coupled MCF (WCMCF) and uncoupled MCF (UCMCF). Their key-parameters are summarized in Tab. 1.

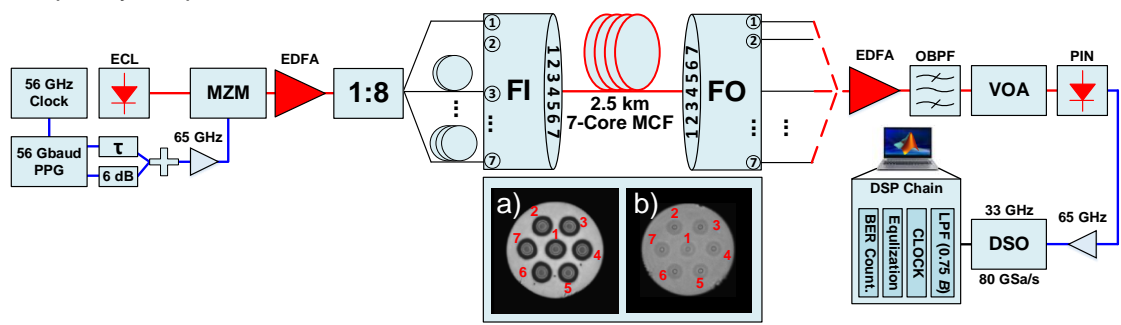

**Fig.1:** Block diagram of the experimental setup. The cross-sections of the used multicore fibers are shown as insets: a) weakly-coupled MCF; b) uncoupled MCF. Acronyms: PPG – pulse pattern generator, ECL – external cavity laser, MZM – Mach-Zehnder modulator, EDFA – Erbium doped fiber amplifier, FI – fan-in, FO – fan-out, OBPF – optical bandpass filter, VOA – variable optical attenuator, DSO – digital storage oscilloscope, DSP – digital signal processing, LPF – lowpass filter.

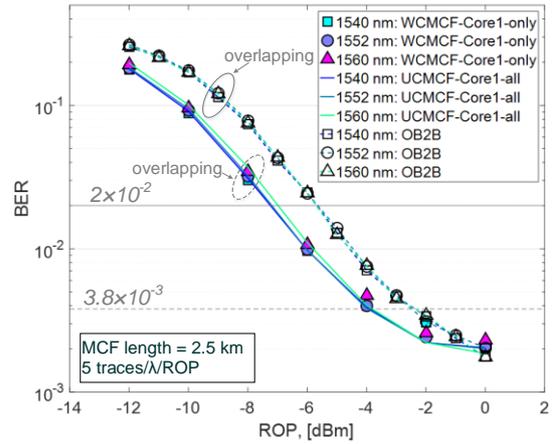

**Fig. 2:** BER curves for the 56-Gbaud PAM-4 showing how an error rate changes with a received optical power and carrier wavelength for the considered transmission scenarios: (1) signals are coupled only into the core 1 of the WCMCF; (2) signals are coupled into all cores of the UCMCF but BERs are measured only in the core 1; (3) optical-back-to-back.

**Tab. 1:** Parameters of the considered 7-core MCF

| Parameter | WCMCF | UCMCF |
|---|---|---|
| Cladding diameter | 150 µm | 150 µm |
| Core pitch | 42 µm | 42 µm |
| Trench | No | Yes |
| Average attenuation | 0.2 dB/km @ 1550 nm | 0.2 dB/km @ 1550 nm |
| Chromatic dispersion | 16 ps/nm/km | 16 ps/nm/km |
| Crosstalk (XT) between adjacent cores | -11dB/ 100 km | -45dB/ 100 km |

After the transmission, the spatial channels are demultiplexed by a fan-out device (FO). Then the signals from each core are sent for a detection. The receiver part consists of an EDFA preamplifier, a wave-shaper, which functioned as an optical bandpass filter (OBPF) and a variable attenuator, a 40 GHz PIN photodiode, a 65 GHz electrical amplifier, and a digital storage oscilloscope (DSO, 33 GHz, 80 GSa/s). Finally, the captured traces of the sampled signals are processed offline with a digital signal processing (DSP) chain, consisting of a low-pass filter, a maximum variance timing recovery, a symbol-spaced decision-feedback equalizer (DFE) with 43 feed-forward taps (FFT) and 12 feedback taps (FBT), and an error counter. Note that the DFE tap weights were obtained for each fiber, each core, and for each wavelength.

**Results and discussion**

This section includes experimental results showing how a bit-error-rate (BER) changes with a received optical power (ROP) along with the BER statistic for each core of the multicore fibers carrying different wavelengths centered to 1540 nm, 1552 nm or 1560 nm with ROPs at the photodetector adjusted to 0 dBm and -1 dBm, respectively.

Figure 2 compares the BER curves for three configurations of the fiber-link: *first*, the 56-Gbaud PAM-4 signals are coupled only into the core 1 of the weakly-coupled multicore fiber while other cores are idle; *second*, signals are coupled into all cores of the uncoupled multicore fiber while the traces of the signals, which are received after transmission over the core 1, are captured by the DSO for further post-processing and a BER counting; *third*, an optical-back-to-back (OB2B). For each data point, five (5) traces are captured and analyzed. Figure 2 displays the mean BER values. It serves as a reference, proving that the difference between the BERs obtained for the WCMCF measurement and for the UCMCF (shown in Fig. 3), is mainly due to obviously higher inter-core crosstalk in the weakly-coupled multicore fiber. If this crosstalk is eliminated, the signal quality would be identical, indicating the similar properties of single core in both types of MCFs. In addition, Fig. 2 proves that the crosstalk levels in the uncoupled MCF are negligibly small to impact on signal quality in a short-reach fiber-optic links employing 56-Gbaud PAM-4 signaling. The BER curves obtained for the OB2B (Fig. 2) show that this experimental setup with its implementation suffers from the chirp-dispersion interaction in bandwidth-limited conditions imposed by the DSO bandwidth. These lead to BER improvement after transmission over the multicore fiber link compared to the OB2B transmission.

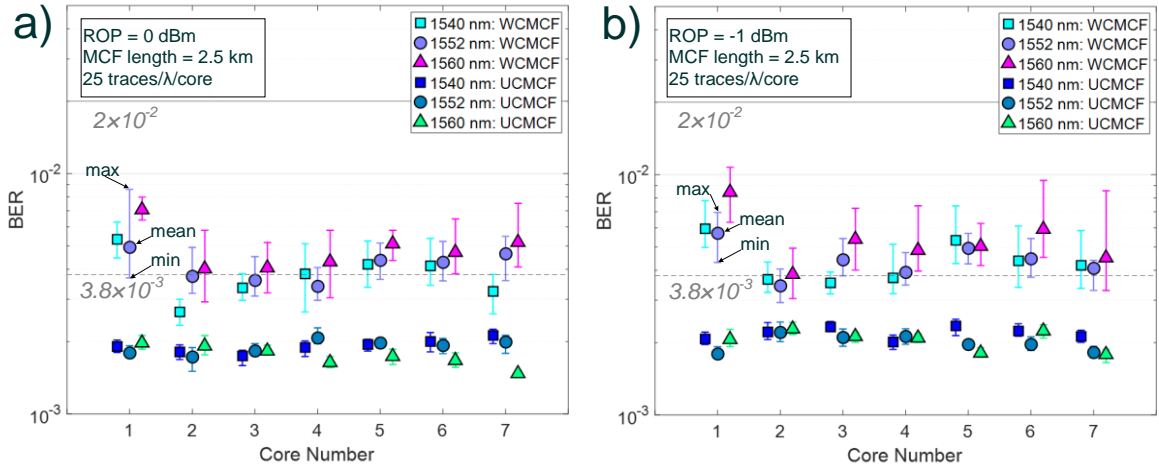

**Fig. 3:** BER statistics accumulated over 1 050 measurements (25 traces/λ/core/fiber) for the 56-Gbaud PAM-4 transmission over the weakly-coupled and uncoupled MCF showing average (illustrated with markers) and range (illustrated with error bars) of BER values detected for each fiber core at two different received optical power values: a) 0 dBm and b) -1 dBm.

Figure 3 illustrates data from 1 050 BER measurements that were performed to reveal the impact of the inter-core crosstalk between adjacent cores on 56-Gbaud PAM-4 transmission over the multicore fibers. These measurements were performed close to the sensitivity threshold of the PD for the hard-decision forward error correction (HD-FEC) of $3.8\times10^{-3}$ to ensure that a sufficient error statistic is accumulated for each captured trace. The results reveal several key-properties of the inter-core crosstalk. *First*, it has a major impact on the received signal quality even after relatively short transmission (like 2.5-km) over the weakly-coupled MCF. *Second*, this crosstalk is wavelength dependent, which means that its impact on signal quality is also wavelength depend. This is confirmed by the BER results. *Third*, the crosstalk dynamic in weakly-coupled MCF, which is theoretically described previously[5], was observed during the measurements and are depicted by the error bars, which can substantially differ from the mean BER values averaged over the number of processed measurements.

**Concluding remarks**

The impact of inter-core crosstalk on a 56-Gbaud PAM-4 transmission over a seven-core fiber is experimentally demonstrated, showing the crosstalk's wavelength dependence as well as revealing its time-dependence in system implementation. The study was performed using a 2.5-km long weakly-coupled and uncoupled multicore fibers having -11 dB/100 km and -45 dB/100km crosstalk between adjacent cores, respectively. The BER results show that shorter wavelengths are more affected by this type of crosstalk in terms of average BER performance, while longer wavelengths have larger fluctuation range, which means that such system with WCMCF has obviously higher instability in time even in short reach systems.

**Acknowledgements**
This work is supported by the Swedish ICT TNG project SCENE, the Swedish Research Council projects PHASE (2016-04510) and Go-iData (2016-04489), the EU H2020-MSCA-IF project NEWMAN (no. 752826), the Swedish Foundation for Strategic Research, Göran Gustafsson Foundation, Natural Science Foundation of Guangdong Province, National Natural Science Foundation of China, and the VINNOVA-funded Celtic-Plus sub-project SENDATE-FICUS. The equipment was funded by Knut and Alice Wallenberg foundation.